\documentclass[12pt]{article}
\usepackage{epsfig}
\makeatletter
\def\alt{\mathrel{\mathpalette\gl@align<}}
\def\agt{\mathrel{\mathpalette\gl@align>}}
\def\gl@align#1#2{\lower.6ex\vbox{\baselineskip\z@skip\lineskip\z@
\ialign{$\m@th#1\hfil##\hfil$\crcr#2\crcr\sim\crcr}}}
\makeatother

\begin{document}
\begin{flushright}
{\tt hep-ph/0210320}\\
OSU-HEP-02-13 \\
October, 2002 \\
\end{flushright}
\vspace*{2cm}
\begin{center}
{\baselineskip 25pt
\large{\bf 
Orbifold Breaking of 3-3-1 Model
}}

\vspace{1cm}

{\large 
Ilia Gogoladze\footnote
{On a leave of absence from: Andronikashvili Institute of Physics, GAS, 380077, Tbilish, Georgia.\\
email: {\tt ilia@hep.phy.okstate.edu}}, 
Yukihiro Mimura\footnote
{email: {\tt mimura@hep.phy.okstate.edu}}
and 
S. Nandi\footnote
{email: {\tt shaown@okstate.edu}}
}
\vspace{.5cm}

{\small {\it Physics Department, Oklahoma State University,
             Stillwater, OK 74078}}

\vspace{.5cm}

\vspace{1.5cm}
{\bf Abstract}
\end{center}

We study the five dimensional $SU(3)_c \times SU(3)_L \times U(1)_X$ gauge theory
(3-3-1 Model) 
in which the gauge symmetry is broken down through orbifold compactification to four dimensions.
The new model has several distinguishing features compared to the usual four dimensional model.
We develop the formalism, and give expressions for the gauge boson masses,
mixings and their couplings to the fermions.
Phenomenology of the model is briefly studied.

\thispagestyle{empty}

\bigskip
\newpage

\addtocounter{page}{-1}

\section{Introduction}
\baselineskip 20pt

The Standard Model (SM) in four dimensions has been well established as an 
effective theory below the weak scale.
There have been many attempts to go beyond the SM.
One attempt is the extension of the gauge group,
while the other being the extension of the number of dimensions beyond the usual
four space-time as motivated by the string theory.

In four dimensions, 
as an extension of the gauge group,
models of $SU(3)\times SU(3)_L \times U(1)_X$
(3-3-1 model) were suggested \cite{Pisano:ee}.
The models include extra gauge bosons, which are called bilepton gauge bosons.
The gauge bosons couple to two leptons, and thus, they have two units of lepton number.
The interesting feature of this model is that the anomaly is not canceled within 
each generation, and that 
the anomaly is canceled among the three generations.
Thus, this model gives an explanation why the number of generation is 3.
Unlike other extended models,
the breaking scale of the $SU(3)_L \times U(1)_X$
have an upper bound about a few TeV
due to the matching condition of the gauge coupling constants.
Hence,
it is very interesting to investigate the phenomenology of this model
and to explore if the extra gauge bosons can be discovered (or ruled out)
in the future high energy colliders such as LHC.

Recent realization that the string scale may be much smaller than the Planck scale,
even as low as few tens of TeV, has inspired study of gauge symmetry in higher dimensions
with the subsequent compactification to four dimension \cite{Witten:1996mz,Arkani-Hamed:1998rs}.
Standard Model and its minimal supersymmetric extension has been formulated in five dimensions
with compactification at an inverse TeV scale and their phenomenological consequences 
have been explored \cite{Pomarol:1998sd,Barbieri:2000vh}.
In these studies, the usual Higgs mechanism has been used to break the gauge symmetry.

Compactification of this higher dimensional theory to four dimensions using a suitable 
manifold or orbifold, and choice of suitable boundary conditions,
opens up a new mechanism for breaking gauge symmetry \cite{Scherk:1978ta}, an interesting alternative to the
usual Higgs mechanism.
A symmetry breaking occurs when the different components in a multiplet of the gauge group $G$
is assigned different quantum number for the discrete symmetry group of the 
compactifying orbifold \cite{Kawamura:1999nj}.
Such orbifold breaking of $SU(5)$ and $SO(10)$ Grand Unified Theories (GUT) and their implications
as well as the symmetry breaking patterns of various group under orbifold compactifications
have been studied \cite{Kawamura:2000ev, Hebecker:2001jb}.
In these studies, orbifold compactifications have been used to break the GUT symmetry to the SM.
Orbifold compactification has also been used to break left-right gauge symmetry group 
giving rise to interesting phenomenological implications \cite{Mimura:2002te}.

In this work, we study the $SU(3)_L \times U(1)_X$ gauge theories in five dimension.
The gauge symmetry is broken to $SU(2)_L \times U(1)_{X^\prime} \times U(1)_X$
upon compactification on a $S^1/Z_2 \times Z_2^\prime$ orbifold.
This is one of the minimal application of the orbifold breaking of a realistic gauge model.
The next stage of symmetry breaking to the SM, $SU(2)_L \times U(1)_Y$ is achieved by using
usual Higgs mechanism.
We assume the gauge and the Higgs bosons propagate into the extra dimensions,
while for the fermions we consider two cases: they can propagate into the bulk 
or they are localized at the 4 dimensional wall (3-brane).
%
There are several interesting differences with the usual four dimensional 3-3-1 model.
The most important qualitative differences are that the bilepton gauge bosons are  
Kaluza-Klein (KK) excited states,
while the extra neutral gauge boson have KK zero mode,
thus the mass of the neutral gauge boson is less than the mass of biletons,
contrary to the 4 dimensional models.
Thus the neutral gauge boson, $Z^\prime$, can have larger contribution to the low energy 
neutral current.
In the usual 4 dimensional 3-3-1 models,
gauge coupling for $U(1)_X$ blows up around 3-4 TeV,
thus we need some new physics around such scale.
The scheme of the TeV scale extra dimension is well compatible to such situation,
since the theory becomes five dimensional beyond this scale.
In the same way to the usual 4 dimensional 3-3-1 models,
we find that the compactification scale has an upper bound around 3-4 TeV.
Thus, these new gauge bosons as well as their KK excitations can be explored at the upcoming LHC.


\section{Formalism: Orbifold breaking of 3-3-1 gauge symmetry}

We start with the 5-dimensional $(x^\mu,y)$ gauge theory
$SU(3)_L \times U(1)_X$ with coupling $g$, $g_X$.
The fifth space-like dimensional coordinate $y$ is compactified in an orbifold
$S^1/Z_2 \times Z_2^\prime$. 
The orbifold is constructed by identifying
the coordinate with $Z_2$ transformation $y \rightarrow -y$
and $Z_2^\prime$ transformation $y^\prime \rightarrow -y^\prime$,
where $y^\prime = y + \pi R/2$.
Then the orbifold space is regarded as a interval $[0, \pi R/2]$
and 4 dimensional wall are placed at the folding point $y=0$ and $y=\pi R/2$.
%
%
%

Using the orbifold compactification,
we break the gauge symmetry to $SU(2)_L \times U(1)_{X^\prime} \times U(1)_X$.
We impose the following transformation property for the five dimensional
$SU(3)_L$ gauge fields under $Z_2 \times Z_2^\prime$. 
\begin{eqnarray}
W_\mu(x^\mu, y) &\rightarrow& W_\mu (x^\mu, -y) = P W_\mu(x^\mu,y) P^{-1}, \\
W_5(x^\mu, y) &\rightarrow& W_5 (x^\mu, -y) = - P W_5(x^\mu,y) P^{-1}, \\
W_\mu(x^\mu, y^\prime) &\rightarrow& W_\mu (x^\mu, -y^\prime) = P^\prime W_\mu(x^\mu,y^\prime) P^{\prime -1}, \\
W_5(x^\mu, y^\prime) &\rightarrow& W_5 (x^\mu, -y^\prime) = - P^\prime W_5(x^\mu,y^\prime) P^{\prime -1}.
\end{eqnarray}
Note that the five dimensional Lagrangian is invariant under the above transformations.
We choose $P= {\rm diag}(1,1,1)$, $P^\prime = {\rm diag}(1,1,-1)$.
For the $U(1)_X$ and $SU(3)_c$ gauge fields, we choose
the $P$ and $P^\prime$ to be identity matrix, so that $U(1)_X$ and $SU(3)_c$ gauge symmetries remain unbroken. 

The five dimensional $SU(3)_L$ gauge filed $W$ can be written as
\begin{equation}
W = \lambda^a W^a = 
\left(
\begin{array}{ccc}
W^3 + \frac1{\sqrt3} W^8 & \sqrt2 W^+ & \sqrt2 Y^{++} \\
\sqrt2 W^- & -W^3 + \frac1{\sqrt3} W^8 & \sqrt2 Y^+ \\
\sqrt2 Y^{--} & \sqrt2 Y^- & -\frac2{\sqrt3} W^8
\end{array}
\right).
\end{equation}
Then, we find that the gauge fields have following parities under $Z_2 \times Z_2^\prime$,
\begin{eqnarray}
W^3_\mu, W^8_\mu, W^\pm_\mu : (+,+) &,& Y^\pm_\mu, Y^{\pm \pm}_\mu : (+,-), \\
W^3_5, W^8_5, W^\pm_5 : (-,-) &,& Y^\pm_5, Y^{\pm \pm}_5 : (-,+).
\end{eqnarray}

The above five dimensional fields can be Fourier expanded as
\begin{eqnarray}
\varphi_{(+,+)}(x^\mu, y) &=& \sqrt{\frac2{\pi R}} \left( \varphi_{(+,+)}^{(0)} (x^\mu)
+ \sqrt2 \sum_{n=1}^{\infty} \varphi_{(+,+)}^{(n)} (x^\mu) \cos \frac{2ny}R \right), 
\label{fourier:W0} \\
\varphi_{(+,-)}(x^\mu, y) &=& 
\frac2{\sqrt{\pi R}} \sum_{n=1}^{\infty} \varphi_{(+,-)}^{(n)} (x^\mu) \cos \frac{(2n-1)y}R, 
\label{fourier:W+} \\
\varphi_{(-,+)}(x^\mu, y) &=& 
\frac2{\sqrt{\pi R}} \sum_{n=1}^{\infty} \varphi_{(-,+)}^{(n)} (x^\mu) \sin \frac{(2n-1)y}R, \\
\varphi_{(-,-)}(x^\mu, y) &=& 
\frac2{\sqrt{\pi R}} \sum_{n=1}^{\infty} \varphi_{(-,-)}^{(n)} (x^\mu) \sin \frac{2ny}R.
\end{eqnarray}
We find the four dimensional fields 
$\varphi^{(n)}_{(+,+)}$ and $\varphi^{(n)}_{(-,-)}$ have masses $2n/R$,
while $\varphi^{(n)}_{(+,-)}$ and $\varphi^{(n)}_{(-,+)}$ have masses $(2n-1)/R$.
Only $\varphi^{(0)}_{(+,+)}$ is massless.
Thus we find that the 4 dimensional bilepton gauge bosons $Y^{\pm}_\mu$, $Y^{\pm \pm}_\mu$ 
do not have massless modes, while $W^3_\mu$, $W^8_\mu$ and $W^\pm_\mu$ have massless modes,
hence, the $SU(3)_L$ symmetry is broken down to $SU(2)_L \times U(1)_{X^\prime}$ symmetry in 4 dimension.

We break the remaining symmetry to the SM gauge symmetry, $SU(2)_L \times U(1)_Y$, by using the Higgs mechanism.
The SM gauge symmetry is also broken down to $U(1)_{EM}$ by Higgs mechanism as usual.
The appropriate Higgs multiplets for this scenario are
\begin{equation}
\Phi: (1,3,1), \quad \Delta: (1,3,0), \quad \Delta^\prime: (1,3,-1), \quad \eta: (1,6,0),
\end{equation}
where the quantum numbers in parenthesis refer to $SU(3)_c \times SU(3)_L \times U(1)_X$ representations.
We define their transformation property under $Z_2 \times Z_2^\prime$ as
\begin{eqnarray}
\Phi (x^\mu, y) &\rightarrow& \Phi(x^\mu, -y) = P \Phi (x^\mu,y), \\
\Phi (x^\mu, y^\prime) &\rightarrow& \Phi(x^\mu, -y^\prime) = -P^\prime \Phi (x^\mu,y), \\
\Delta (x^\mu, y) &\rightarrow& \Delta(x^\mu, -y) = P \Delta(x^\mu, y) , \\
\Delta (x^\mu, y^\prime) &\rightarrow& \Delta(x^\mu, -y^\prime) = P^\prime \Delta(x^\mu, y^\prime) , \\
\Delta^\prime (x^\mu, y) &\rightarrow& \Delta^\prime(x^\mu, -y) = P \Delta^\prime(x^\mu, y) , \\
\Delta^\prime (x^\mu, y^\prime) &\rightarrow& \Delta^\prime(x^\mu, -y^\prime) = P^\prime \Delta^\prime(x^\mu, y^\prime) , \\
\eta (x^\mu, y) &\rightarrow& \eta(x^\mu, -y) = P\eta(x^\mu, y) P^{-1}, \\
\eta (x^\mu, y^\prime) &\rightarrow& \eta(x^\mu, -y^\prime) = -P^\prime \eta(x^\mu, y^\prime) P^{\prime -1}.
\end{eqnarray}
The $SU(3)_L$ triplets and sextet are decomposed to $SU(2)_L\times U(1)_{X^\prime}$
as $3= (2,1)+(1,-2)$ and $6 = (3,2)+(2,-1)+(1,-4)$.
The decomposed fields have the following $Z_2\times Z_2^\prime$ transformation property:
\begin{eqnarray}
\Phi_1, \Delta_2, \Delta^\prime_2, \eta_2 &:& (+,+) \\
\Phi_2, \Delta_1, \Delta^\prime_1, \eta_3, \eta_1 &:& (+,-)
\end{eqnarray}
where the suffix denotes $SU(2)_L$ representations.
Thus, while $\Phi_2 = (\phi^{++},\phi^+)$ doublet acquires a KK mass of $O(1/R)$,
$\Phi_1 = \phi^0$ (which electric charge is zero) have a massless mode,
and thus can be assigned a vacuum expectation value (VEV) (we call its VEV $u$)
to break the remaining gauge symmetry down to SM gauge symmetry. 
All the $SU(2)_L$ doublets is chosen to have VEVs in order to break SM gauge group
and to give appropriate masses to fermions.
If we denote the VEVs of neutral component of $SU(2)_L$ doublets, $\Delta_2$, $\Delta^\prime_2$ and $\eta_2$
as $v$, $v^\prime$ and $w$ respectively,
the scale of the VEVs are assumed to be
\begin{equation}
u \gg v, v^\prime, w \simeq O(100) \ {\rm GeV}.
\end{equation}

The sextet $\eta$ is necessary in order to obtain acceptable masses for charged lepton
in the case electron, positron and left-handed neutrino is in one $SU(3)_L$ triplet.
The VEV in the neutral component of $\eta_3$ gives a large majorana mass to left-handed neutrino,
so that the VEV should be zero or extremely small.
In the four dimensional theory, a fine tuning in the Higgs potential is needed to achieve this.
In our model, if we choose the doublet $\eta_2$ has a massless mode,
$\eta_3$ automatically does not have a massless mode, and thus its neutral component will not have
a VEV.
This is 
a very nice feature of our orbifold breaking scenario.

\section{Gauge Boson Masses}

The Frampton-Pleitez type 3-3-1 models have an upper bound for extra gauge bosons
due to the evolution of gauge couplings.
To see that, we first study matching condition of gauge couplings.

The $SU(3)_L$ gauge symmetry is broken down to $SU(2)_L \times U(1)_{X^\prime}$ 
by orbifold compactification.
The VEV of Higgs $\Phi$ breaks remaining $U(1)_{X^\prime} \times U(1)_X$ symmetry
down to $U(1)_Y$ symmetry.
Then, the hypercharge is defined as
\begin{equation}
Y = 2 (\sqrt3 T^8 +  X I),
\end{equation}
and electric charge Q is defined as
\begin{equation}
Q =  T^3 + \frac{Y}2.
\end{equation}
Here $T^a$ is the $SU(3)_L$ generator normalized as
Tr $(T^a T^b) = 1/2$ $\delta^{ab}$,
and $I =$ diag$(1,1,1)$.
We define the covariant derivative for $SU(3)_L$ triplets as
\begin{equation}
D_\mu = \partial_\mu - ig T^a W_\mu^a - ig_X X I V_\mu.
\end{equation}
The $W^a_\mu$ and $V_\mu$ are the $SU(3)_L \times U(1)_X$ gauge fields.
At this normalization,
the coupling constant of $U(1)_Y$, $g_Y$, is given as
\begin{equation}
\frac1{g_Y^2} = \frac1{g_{X^\prime}^2} + \frac1{g_X^2}.
\end{equation}
where $g_{X^\prime}$ is a coupling constant for $U(1)_{X^\prime}$.
At the unification scale for $SU(3)_L$,
we have the unification condition
\begin{equation}
g_{X^\prime} = \frac1{\sqrt3} g.
\end{equation}
Since the $SU(3)_L$ gauge symmetry is broken on the brane at $y=\pi R/2$
in our model, we can add the brane gauge interaction which breaks the 
unification condition of gauge couplings.
However, we can neglect the correction of brane interaction 
if the cutoff scale is much larger than compactification scale \cite{Hebecker:2001jb}.
{}From this matching condition,
we obtain the relation at the compactification scale $M_c$,
\begin{equation}
\frac{g_X^2}{g^2} = \frac{\sin^2 \theta_W(M_c)}{1-4 \sin^2 \theta_W(M_c)},
\end{equation}
where the weak mixing angle is defined as $\tan \theta_W = g_Y/g$.
Therefore, $\sin^2 \theta_W (M_c)$ has to be less than 1/4.
This constraint $\sin^2\theta_W (M_c) < 1/4$ demands an upper bound of the compactification scale.
If we demands that the gauge coupling $g_X$ at $M_c$ is less than $N$,
we obtain the constraint as
\begin{equation}
\frac{1- 4 \sin^2 \theta_W(M_c)}{\alpha_{EM} (M_c)} > \frac1N.
\end{equation}
Some of the fermion such as new exotic quarks of this models decouple
at the breaking scale of $U(1)_X \times U(1)_{X^\prime}$.
We call the scale $M_D$.
Assuming that $M_c > M_D$, we obtain that the compactification scale is less than 4.4 TeV.
We used the experimental data, $\alpha_{EM}^{-1} (M_Z) = 127.9$ and $\sin^2\theta_W= 0.2311$ 
in the $\overline{\rm MS}$ scheme.
We plot the upper bound of $M_c$ as a function of $M_D$ in the Fig.1.

\begin{figure}
\begin{center}
\epsfig{file=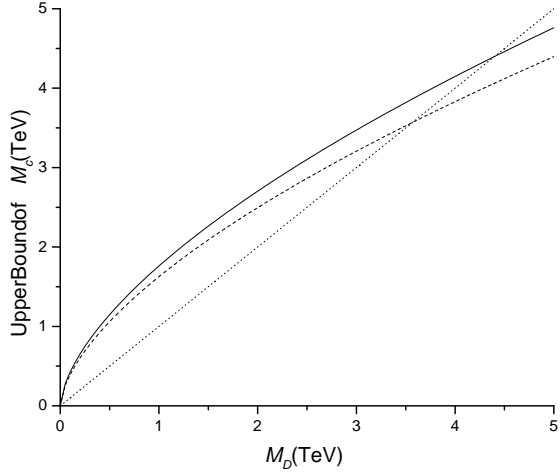,width=8cm} 
\end{center}
\caption{Upper bound for compactification scale $M_c$.
Solid line for $\alpha_X(M_c) < \infty$. Dashed line for $\alpha_X(M_c) < 4\pi$.
Dotted line for reference of $M_c= M_D$. }
\label{fig1}
\end{figure}

Let us see the mass spectra of this model.
We denote the VEVs of the neutral component of Higgs fields,
$\Phi_1$, $\Delta_2$, $\Delta^\prime_2$, and $\eta_2$
as $u/\sqrt2$, $v/\sqrt2$, $v^\prime/\sqrt2$, and $w/\sqrt2$,
respectively.
Then charged gauged bosons acquire masses
\begin{eqnarray}
M_{W_{(n)}}^2 &=& \frac14 g^2 (v^2 + v^{\prime 2} + w^2) + \left(\frac{2n}R \right)^2  \quad (n=0,1,2,\cdots),  \\
M_{Y_{(n)}^+}^2 &=& \frac14 g^2 (u^2 + v^2 +w^2) +\left(\frac{2n-1}R \right)^2 \quad (n=1,2,3, \cdots), \label{eq32}\\
M_{Y_{(n)}^{++}}^2 &=& \frac14 g^2 (u^2 + v^{\prime 2} +w^2) +\left(\frac{2n-1}R \right)^2 \quad (n=1,2,3, \cdots).\label{eq33}
\end{eqnarray}
For neutral gauge bosons, we can identify the photon field $\gamma$ and the massive bosons
$Z$ and $Z^\prime$ as follows.
\begin{eqnarray}
\gamma &=& \sin\theta_W W^3 + \cos\theta_W (\sqrt3 \tan\theta_W W^8 + \sqrt{1-3\tan^2\theta_W} V), \\
Z &=& \cos\theta_W W^3 - \sin\theta_W (\sqrt3 \tan\theta_W W^8 + \sqrt{1-3\tan^2\theta_W} V), \\
Z^\prime &=& - \sqrt{1-3\tan^2\theta_W} W^8 + \sqrt3 \tan\theta_W V.
\end{eqnarray}
In this basis, the mass squared matrix for $Z_{(n)}$ and $Z^\prime_{(n)}$ $(n=0,1,2, \cdots)$ is given by
\begin{equation}
{\cal M}^2_{(n)} = \left(
\begin{array}{cc}
M_Z^2 & M_{ZZ^\prime}^2 \\
M_{ZZ^\prime}^2 & M_{Z^\prime}^2
\end{array}
\right),
\end{equation}
with
\begin{eqnarray}
M_Z^2 &=& \frac14 \frac{g^2}{\cos^2 \theta_W} (v^2 + v^{\prime 2} + w^2) + \left(\frac{2n}R \right)^2, \\
M_{Z^\prime}^2 &=& \frac13 g^2 \left( 
\frac{\cos^2\theta_W}{1-4\sin^2 \theta_W} u^2
+ \frac{1-4\sin^2 \theta_W}{4\cos^2 \theta_W} (v^2 + v^{\prime2} +w^2)
+ \frac{3\sin^2 \theta_W}{1-4 \sin^2 \theta_W} v^{\prime2}
\right) + \left(\frac{2n}R \right)^2, \nonumber \\
\\
M_{ZZ^\prime}^2 &=& \frac1{4\sqrt3} g^2 \frac{\sqrt{1-4\sin^2\theta_W}}{\cos^2\theta_W}
\left(
v^2 + w^2 - \frac{1+2 \sin^2\theta_W}{1-4\sin^2\theta_W} v^{\prime2}
\right).
\end{eqnarray}
The mass eigenstates $Z_1$ and $Z_2$ are
\begin{eqnarray}
Z_1 &=& \cos \phi Z - \sin \phi Z^\prime, \\
Z_2 &=& \sin \phi Z + \cos \phi Z^\prime,
\end{eqnarray}
where the mixing angle $\phi$ is given by
\begin{equation}
\tan 2\phi = \frac{2M_{ZZ^\prime}^2}{M_{Z^\prime}^2-M_Z^2}.
\end{equation}
Assuming that $u> v,v^\prime,w$, 
we obtain
\begin{equation}
M_{Z_1}^2 \simeq M_Z^2 (1- \frac{M_{Z^\prime}^2}{M_Z^2} \phi^2), \qquad
\phi \simeq \frac{M_{ZZ\prime}^2}{M_{Z\prime}^2}.
\end{equation}
The mixing angle have a range
\begin{equation}
-\frac{1+2 \sin^2\theta_W}{\sqrt3 \sqrt{1-4\sin^2\theta_W}} \frac{M_{Z_1}^2}{M_{Z_2}^2}
\leq \phi \leq 
\frac{\sqrt{1-4\sin^2\theta_W}}{\sqrt3} \frac{M_{Z_1}^2}{M_{Z_2}^2}.
\label{range_of_phi}
\end{equation}
Note that while the bilepton gauge bosons, $Y^{\pm\pm}$ and $Y^\pm$, acquire masses at 
the compactification scale (see Eq. (\ref{eq32}-\ref{eq33}),
the extra neutral gauge boson, $Z_2$ does not. Thus, $M_{Z_2} \ll M_Y$.
This is exactly opposite to the usual Higgs breaking of the four dimensional model
in which $M_{Z_2} \gg M_Y$.
Thus orbifold breaking naturally predicts a low scale extra $Z^\prime$ boson 
giving rise to interesting phenomenological implications as we will discuss later.

\section{Fermions}

In our model, the $SU(3)_L$ gauge symmetry in five dimension
is broken by the boundary conditions.
At the boundary, $y=\pi R/2$, the five dimensional gauge fields
$Y^\pm$ and $Y^{\pm\pm}$ vanish.
Thus, the orbifold fixed point
$y=\pi R/2$ does not respect the $SU(3)_L$ symmetry.
When we include the fermions in the theory,
we have two choices: the fermions can propagate
into the bulk or they are localized at the 
4D wall orbifold fixed points.
%
%
%

We show two different anomaly-free fermion representations.
The first anomaly free fermion contents are following \cite{Pisano:ee}.
\begin{equation}
\ell_a = (e_a,\nu_a,e^c_a)^T : (1,3^*,0), 
\end{equation}
\begin{equation}
q_i = (u_i, d_i, D_i)^T : (3,3,-\frac13), \quad
q_3 = (d_3, u_3, T)^T : (3,3^*,\frac23), 
\end{equation}
\begin{equation}
u^c_a : (3^*,1,-\frac23), \quad
d^c_a : (3^*,1,\frac13), \qquad
D_i^c : (3^*,1,\frac43), \quad
T^c   : (3^*,1,-\frac53),
\end{equation}
where $a=1,2,3$ is a family index and $i=1,2$ is related to two of the three families.

The $D_i$ and $T$ are additional quark which have exotic electric charges,
$-4/3$ and $5/3$, respectively.
In this model, the anomaly is not canceled within each generation, but canceled
among three generations
by choosing that two of the three generations are $(3,3,-1/3)$ and
the other one generation is $(3,3^*,2/3)$.
There is an ambiguity which generation we assign the $(3,3^*,2/3)$ to.

Since both electron and positron are in the same $SU(3)_L$ triplet,
we need Higgs sextet to give the physically acceptable masses to charged leptons:
otherwise the mass matrix becomes anti-symmetric giving rise to $m_e=0$ and $m_\mu=m_\tau$.
We can consider other assignment for positron,
that is, the positron is additional $SU(3)_L$ singlet $(1,1,1)$ and $e^c$ component in the triplet $\ell$
get large by adding $E:(1,1,-1)$.
At this choice, we denote the multiplets as
\begin{equation}
\ell_a = (\nu_a,e_a,E^c_a)^T : (1,3^*,0), \quad e^c_a : (1,1,1), \quad E_a : (1,1,-1).
\end{equation}

We can have another type of anomaly-free fermion set.
\begin{equation}
\ell_a = (\nu_a,e_a,E^{--}_a)^T : (1,3,-1), \quad e^c_a : (1,1,1), \quad E^{++}_a : (1,1,2),
\end{equation}
\begin{equation}
q_i = (d_i, u_i, U_i)^T : (3,3^*,\frac23), \quad
q_3 = (u_3, d_3, B)^T : (3,3,-\frac13), 
\end{equation}
\begin{equation}
 u^c_a : (3^*,1,-\frac23), \quad
 d^c_a : (3^*,1,\frac13), \qquad
 U_i^c : (3^*,1,\frac43), \quad
 B^c   : (3^*,1,-\frac53).
\end{equation}

Let us see the case that fermions are in the bulk.
The bulk fermions have both chirality in the sense of four dimensional Weyl fermions.
For example, for the lepton triplet, $\ell = (e,\nu,e^c)^T : (1,3^*,0)$,
the fermions have $Z_2 \times Z_2^\prime$ charge as
\begin{equation}
(e,\nu)_L : (++) , \quad e^c_L : (+-), \quad (e,\nu)_R : (--), \quad e^c_R : (-+).
\end{equation}
Only $(e,\nu)_L$ has zero mode.
We have to have one more lepton triplet such as 
\begin{equation}
(e^\prime,\nu^\prime)_L : (+-) , \quad e^{c\prime}_L : (++), \quad (e^\prime,\nu^\prime)_R : (-+), \quad e^{c\prime}_R : (--),
\end{equation}
and there is a positron in the triplet.
Since electron and positron are in the different multiplet in this case,
we don't need Higgs sextet to construct a phenomenological viable charged lepton mass matrix.

For the quark fields, we also need to make them doubled in the same way
in order to cancel the brane-localized chiral anomaly \cite{Arkani-Hamed:2001is}.


The most interesting feature of the bulk-fermion case, the lightest KK particles (LKP)
become a candidate of dark matter,
since the LKP become stable if all the particles are bulk fields.

\section{Phenomenological Implications}

In original 3-3-1 model, the VEV of Higgs field $\Phi$ breaks 
$SU(3)_L \times U(1)_X$ down to SM gauge group directly.
Thus, the mass of the neutral extra gauge boson, $Z^\prime$, 
is related to the masses of new charged gauge bosons, $Y^{\pm}$ and $Y^{\pm\pm}$.
The charged gauge bosons' masses are constrained from
collider experiments and muon decay.
The constraint of the charged gauge boson pushes the lower bound of $Z^\prime$ boson mass
to be larger than $O(1)$ TeV.
In the 5 dimensional 3-3-1 model, however,
the $SU(3)_L \times U(1)_X$ symmetry cannot be broken down to SM directly.
Therefore, there is no relation between the masses of new charged and neutral gauge bosons.
As a result, the mass of $Z^\prime$ can be lower than 1 TeV,
and hence its contribution to the low energy experiments can be significant.

To see that, we first write out the interaction of $Z^\prime$ boson and fermions.
The gauge interaction is given as
\begin{equation}
{\cal L}(Z^\prime) = \frac{g}{\cos\theta_W} Z^{\prime \mu} 
\bar{f} \gamma_\mu \left(
g_L^\prime (f) \frac{1-\gamma_5}2 + g_R^\prime (f) \frac{1+\gamma_5}2 
\right) f.
\end{equation}
The coupling coefficients are given as
\begin{equation}
g^\prime_{L,R} (f) = -\frac{\sqrt{1-4\sin^2\theta_W}}{2\sqrt3} Y(f_{L,R})
                  + \frac{1-\sin^2\theta_W}{\sqrt3\sqrt{1-4\sin^2\theta_W}} X(f_{L,R}).
\end{equation}
The $Z^\prime$ coupling to the fermion with non-zero $X$ charge is enhanced
since the weak mixing angle is close to 1/4.

In fact, experimental data for forward-backward asymmetry $A_{FB}^{(0,b)}$
and atomic parity violation have more than 2$\sigma$ discrepancy compared 
with SM prediction.
If we assign the $X$ charge of lepton multiplets to be zero,
those predictions might be adjusted
without changing the leptonic experimental data much.

We first concentrate our discussion to atomic parity violation.
The value of the atomic parity violation for Cs is 
\begin{eqnarray}
Q_W (\mbox{SM prediction}) &=& -73.09 \pm 0.03, \\
Q_W (\mbox{experiment})    &=& -72.06 \pm 0.28 \pm 0.34.
\end{eqnarray}
We find the contribution of this 3-3-1 model as
\begin{eqnarray}
\delta Q_W &\simeq&  157 \frac{M_{Z_1}^2}{M_{Z_2}^2} - 980 \phi - 73 \phi^2 \frac{M_{Z_2}^2}{M_{Z_1}^2} 
\quad (X_{Q_1} = \frac23), \\
\delta Q_W &\simeq&  -45 \frac{M_{Z_1}^2}{M_{Z_2}^2} - 300 \phi - 73 \phi^2 \frac{M_{Z_2}^2}{M_{Z_1}^2} 
\quad (X_{Q_1} = -\frac13),
\end{eqnarray}
where $X_{Q_1}$ is the $X$ charge of the first generation of left-handed quarks.
We can adjust the experimental data for the range
$M_{Z_2} \simeq $ 1 TeV and $|\phi| \alt 10^{-3}$.



Next we consider the nucleon-neutrino scattering.
The contribution of $g_L^2$ and $g_R^2$ are following.
For the case that $X$ charge of lepton triplets, $X_\ell$, is 0,
\begin{eqnarray}
\delta g_L^2 &\simeq& -0.025 \frac{M_{Z_1}^2}{M_{Z_2}^2} + 0.0659 \phi + 0.59 \phi^2 \frac{M_{Z_2}^2}{M_{Z_1}^2} 
\quad (X_{Q_1}=\frac23), \\
\delta g_L^2 &\simeq& 0.013 \frac{M_{Z_1}^2}{M_{Z_2}^2} - 0.18 \phi + 0.59 \phi^2 \frac{M_{Z_2}^2}{M_{Z_1}^2} 
\quad (X_{Q_1}=-\frac13), \\
\delta g_R^2 &\simeq& -0.059 \frac{M_{Z_1}^2}{M_{Z_2}^2} + 0.36 \phi + 0.058 \phi^2 \frac{M_{Z_2}^2}{M_{Z_1}^2}.
\end{eqnarray}
This correction is not enough to adjust NuTeV results \cite{Zeller:2001hh}
since we have a constraint of the $Z$-$Z^\prime$ mixing angle $|\phi| \alt 10^{-3}$ from other experiments \cite{Pisano:ee}.

For the case $X_\ell = -1$,
\begin{eqnarray}
\delta g_L^2 &\simeq& 0.49 \frac{M_{Z_1}^2}{M_{Z_2}^2} + 1.9 \phi + 0.59 \phi^2 \frac{M_{Z_2}^2}{M_{Z_1}^2} 
\quad (X_{Q_1}=\frac23), \\
\delta g_L^2 &\simeq& -0.26 \frac{M_{Z_1}^2}{M_{Z_2}^2} + 1.7 \phi + 0.59 \phi^2 \frac{M_{Z_2}^2}{M_{Z_1}^2} 
\quad (X_{Q_1}=-\frac13), \\
\delta g_R^2 &\simeq& 1.14 \frac{M_{Z_1}^2}{M_{Z_2}^2} + 0.5 \phi + 0.058 \phi^2 \frac{M_{Z_2}^2}{M_{Z_1}^2}.
\end{eqnarray}
In this case, we can adjust the NuTeV experiment for the range $M_{Z_2} \simeq 3$ TeV
and $|\phi| \simeq 10^{-3}$ with $\phi<0$.

We now briefly discuss the collider signals for this model.
Since the $Z^\prime$ remains massless after the orbifold breaking 
of the symmetry,
this scenario predicts a low scale $Z^\prime$.
In the original four dimensional model,
$Z^\prime$ obtains mass at a scale higher than that of $Y$
and thus $M_{Z^\prime} \gg M_Y$.
The exactly opposite is true in our orbifold breaking
scenario, and unlike the usual four dimensional model,
the mass of $Z^\prime$ is not constrained from the low energy limit on the mass of $Y$.
Such a low mass $Z^\prime$, probably around a TeV or less,
is likely to show up at the Tevatron Run2 and
will be easily observed at the LHC.
Another interesting signal is the production of $Y^{++}$
in association with exotic charged quark $D$ in the hadronic collider
via the process
$u+g \rightarrow Y^{++} +D$ \cite{Dutta:1994pd}.
$Y^{++}$ will decay to $\ell^+\ell^+$ $(\ell= e,\mu,\tau)$, while
$D$ will decay to $u \ell^-\ell^-$.
This will lead to a spectacular signal of four charged leptons
in the final state with a peak in $\ell^+\ell^+$ combination.
Such a signal will be easily observed at the LHC for the mass range
of $Y$ and $D$ predicted by our scenario.

\section{Conclusions}

We conclude summarizing our main results.
We have considered the orbifold breaking of the 3-3-1 model in the five dimension.
The first stage of symmetry breaking, $\it i.e.$
$SU(3)_L$ gauge symmetry to $SU(2)_L \times U(1)_{X^\prime}$
is achieved via orbifold compactification
to four dimension.
The model has several distinguishing features from the usual four dimensional 3-3-1 model.

First, in the context of the four dimensional 3-3-1 model, we expect new physics
since the gauge coupling of $U(1)_X$ blows up around 4 TeV.
TeV scale compactification is a good candidate as the new physics.
In other words,
this model is very compatible with TeV scale unification
since it predicts that $\sin^2 \theta_W$ is less than 1/4 at
unification scale.

Next, orbifold construction of 3-3-1 model has an advantage to forbid the VEV of $SU(2)_L$ triplet.
In the minimal choice of the fermion representation,
the lepton doublet and positron are included in one $SU(3)_L$ triplet $\ell$.
In that case, we need the sextet Higgs, $\eta$, to give phenomenologically viable mass matrix of charged lepton.
Then the Yukawa coupling $\ell \ell \eta$ contains the majorana mass of neutrino,
if the $SU(2)_L$ triplet in $\eta$ acquire VEV.
Thus, we have to choose the Higgs potential to prohibit the unwanted VEV.
Orbifold compactification gives mass of $1/R$ to the $SU(2)_L$ triplet component naturally,
if the sextet $\eta$ is a bulk Higgs.

Finally, phenomenological point of view,
the mass of extra neutral gauge boson $Z^\prime$ can be lower than 1 TeV in the orbifold breaking scenario.
If the gauge symmetry $SU(3)_L \times U(1)_X$ is directly broken to SM gauge group,
there is a rigid relation between the mass of bilepton gauge bosons and extra neutral gauge boson.
We have a constraint for the mass of bileptons from muon decay or other experiments,
which then lead to a
constraint that the mass of $Z^\prime$ should be larger than 1.3 TeV.
However, in the orbifold breaking scenario,
we have two independent brekaing scale, that is compactification scale and the VEV of triplet Higgs,
and thus, there is no relation between mass of the bileptons and $Z^\prime$.
In that case, the mass of $Z^\prime$ can be lower than 1 TeV,
and is likely to be observed at the high energy collider such as Tevatron Run2 or LHC.
Such a $Z^\prime$ can also contribute the low energy neutral current experiments.
In fact, this model gives interesting contribution to the atomic parity violation
and nucleon-neutrino scattering.

\section*{Acknowledgments}

We thank K.S. Babu for useful discussions.
I.G.and S.N. acknowledge the warm hospitality and support of the 
Fermilab Theory Group from their Summer Visitor Program
during the initial stage of this work.
S.N. also acknowledges the warm hospitality and support of the DESY and 
CERN Theory Groups during his summer visits there.
This work was supported in part by US DOE Grants \# DE-FG030-98ER-41076
and DE-FG-02-01ER-45684.

\end{document}